\documentclass[twocolumn,
aps,nofootinbib,showpacs,showkeys,preprint
tightenlines,preprintnumbers,] {revtex4}

\usepackage{epsf,epsfig,subfigure,graphicx,amsmath,amssymb}
\usepackage{color}
\newcommand{\dis}[1]{\begin{equation}\begin{split}#1\end{split}\end{equation}}
\newcommand{\VEV}[1]{\langle #1 \rangle}
\newcommand{\OVER}[1]{\,\overline{\hskip -0.5mm #1}}

\newcommand{\etal}{{\it et al.}}

\newcommand{\Mp}{M_P}
\newcommand{\tev}{\,\textrm{TeV}}
\newcommand{\gev}{\,\textrm{GeV}}

\newcommand{\kev}{\,\textrm{keV}}
\newcommand{\eV}{\,\textrm{eV}}

\def\Xq{{\cal X}}

\def\MGUT{M_{{\cal X}^{(\rm G)}}}
\def\MI{M_{\rm int}}
\def\phiq{{\phi_{\rm de}}}

\def\one{{\bf 1}}

\def\three{{\bf 3}}

\newcommand{\Z}{{\bf Z}}

\newcommand{\ie}{{\it i.e.~}}

\newcommand{\Xz}{{X^{(0)}}}
\newcommand{\Xzc}{{\overline{X}^{\,(0)}}}

\newcommand{\Xqz}{{{\cal X}^{(0)}}}
\newcommand{\Xqzc}{{\,\overline{\hskip -0.5mm \cal X}^{\,(0)}}}

\newcommand{\Mint}{{ M_{\rm int} }}
\newcommand{\FDE}{{ F_{\rm DE} }}
\newcommand{\Vde}{{$10^{-46}\,\gev^{\,4}$}}

\newcommand{\Quint}{{\it quintessence-flavor-democracy}}
\newcommand{\UDE}{{U(1)$_{\rm de}$}}
\newcommand{\UPQ}{{U(1)$_{\rm PQ}$}}

\begin{document}
\draft

\title{\Large\bf Modeling small dark energy scale with\\ quintessential pseudoscalar boson
}

\author{Jihn E.  Kim }
\affiliation
{
 Department of Physics, Kyung Hee University, Seoul 130-701, Republic of Korea
}

\begin{abstract}
Democracy among the same type of particles is a useful paradigm in studying masses and interactions of particles with supersymmetry(SUSY) or without SUSY. This simple idea predicts the presence of massless particles.
We attempt to use one of these massless pseudoscalar particles as generating the cosmological dark energy(DE) potential. To achieve
the extremely shallow potential of DE, the pseudoscalar boson is required not to couple to the QCD anomaly. So, we consider two pseudoscalars, one coupling to the QCD anomaly (\ie QCD axion) and the other not coupling to the QCD anomaly. To obtain these two pseudoscalars, we introduce two approximate global U(1) symmetries to realize them as the pseudo-Goldstone bosons of the spontaneously broken U(1)'s. These global symmetries are dictated by a gravity respected discrete symmetry. Specifically, we consider an $S_2(L)\times S_2(R)\times \Z_{10R} $ example, and attempt to obtain the DE scale in terms of two observed fundamental mass scales, the grand unification scale $M_G$ and the electroweak scale $v_{\rm ew} $.

\end{abstract}


\maketitle

\section{Introduction}

Massless particles around the electroweak(EW) scale determine physical laws observed around us.
The recent discovery of the Higgs boson \cite{Higgs126} almost confirms the standard model(SM) spectrum below about 1 TeV.
Obtaining such a massless spectrum theoretically has a long history under the name of the `gauge hierarchy problem' \cite{Gildener76}.
Since quarks and leptons have the chiral structure, they can be kept massless down to the TeV scale. However, the Higgs
boson is difficult to be kept massless in the SM framework. The most studied solution of the gauge hierarchy problem has been
supersymmetry(SUSY). Even in the minimally supersymmetrized SM (MSSM), there is a problem known as the `$\mu$ problem' \cite{KimNilles84}. The MSSM $\mu$ term, $\mu H_u H_d$, is required to be at the electroweak scale, and the difficulty obtaining it at the TeV scale is the problem. Recently, the permutation discrete symmetry $S_2\times S_2$ has been used to obtain a massless pair of Higgs doublets, $H_u$ and $H_d$ \cite{Kim13}.

Using a discrete symmetry to obtain the $\mu$ term is welcome in view of the gravity effects violating some ad hoc symmetries
except the gauge symmetry \cite{DiscrGauge89}. Indeed, it was argued in \cite{GravityGlobal} that the Peccei-Quinn(PQ) global
symmetry \cite{PQ77, InvAxionRev10} broken at the intermediate scale has a severe fine tuning problem due to the gravity violation of the global symmetry. If we consider a global symmetry, it must be an approximate symmetry. Discrete symmetries cannot escape this gravity conundrum, but if it is a discrete subgroup of a gauge symmetry or dictated by string theory then it is safe from the gravity conundrum. If the exact discrete symmetries are known, they can be helpful in obtaining an approximate global symmetry.
Therefore, theoretical consideration of a global symmetry proceeds as follows:

\begin{itemize}
\item[1)]
Firstly, at the energy scale much below $M_P$, a global symmetry respected by the $d=2,3, \cdots,d_M$ superpotential terms can be considered. Since it is an approximate symmetry, it must be violated by higher order terms beyond $d_M$ suppressed by the Planck scale or the GUT scale masses.
\item[2)]
Second, the global symmetry in consideration from $d=2,3, \cdots,d_M$ superpotential terms must be a part of a discrete symmetry respected by gravity so that the $d=2,3, \cdots,d_M$ superpotential terms are not affected by the wormhole effects.
\item[3)]
Third, for a specific global symmetry definition, all the superpotential terms up to $d=d_M$ may be needed for a unique definition.
\end{itemize}
The second and third steps are strong conditions. This idea was presented in Ref. \cite{KimNilles13} and here we realize the scheme explicitly. The regions for the discrete and global symmetries are shown in Fig. 2 of \cite{KimNilles13}. Also, there is a common region which defines the approximate global symmetry from the discrete symmetry origin. If we consider the SM gauge group below the GUT scale, the rank of gauge symmetries we can count on are limited in the ultraviolet completed theory. Therefore, if there are too many singlets needed for the concrete definition of the global symmetry, the global symmetry may not be allowed from the ultraviolet completed theory. In Ref. \cite{Kim13},
by obtaining a TeV scale $\mu$ the approximate PQ symmetry toward a strong CP solution was obtained.

The most curious cosmological observation is the extremely small dark endrgy(DE) of order $(0.003\,\eV)^4$ \cite{DE98}. The simplest form of DE is the so-called cosmological constant(CC). The CC appears as a real constant in the Einstein equation \cite{Einstein17} and behaves as energy. On the other hand, if time-dependent homogeneous energy density evolves very slowly and stays almost constant during our observable time scale, it is called DE. So, DE includes the CC. At present, it is fair to say that the CC problem is not understood  \cite{WeinbergCCrev89}. As in any attempts to derive a small DE scale \cite{Quintessence}, here we assume that the CC is zero. The anthropic bound \cite{Weinberg87} is not a calculational scheme.

\begin{figure}[!t]
\begin{center}
\includegraphics[width=0.95\linewidth]{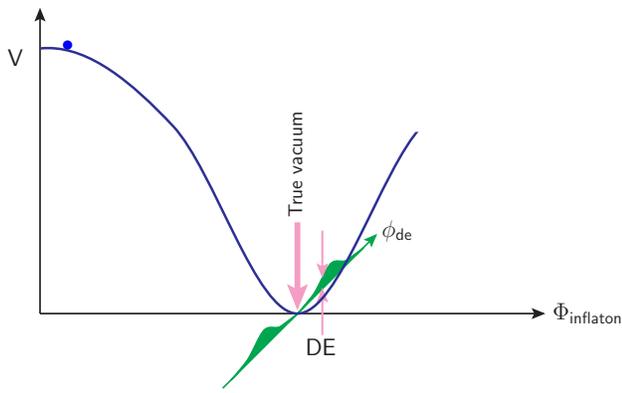}
\end{center}
\caption{A cartoon for the inflation and DEPS potentials shown as the blue and green curves, respectively. After inflation, DE may not be vanishing as shown with the height of the green band. The height of the green band is exaggerated roughly by a factor of $10^{115}$. At the point the thick arrow directs to, all equations of motion are satisfied and the CC is vanishing.} \label{fig:Inflation}
\end{figure}

The almost flat green-colored cartoon potential of Fig. \ref{fig:Inflation} is assumed to be that of a quintessential pseudoscalar field $\phiq$ contributing to DE.
It is different from the earlier terminology `quintessential axion' discussed in \cite{quintAx} where the non-abelian anomaly breaks the global symmetry. The pseudoscalar we consider in this paper is assumed not to couple to a non-Abelian anomaly. Since the terminology `axion' is too much connected to the anomaly, in this paper we prefer to use the name `$\underline{\rm DE}~\underline{\rm p}$eudo$\underline{\rm s}$calar boson'(DEPS) which does not couple to the QCD (and hidden-sector if present) non-Abelian anomalies. The corresponding approximate global symmetry is denoted as \UDE.

The easiest way to introduce the DEPS scale is just assuming a spontaneously broken \UDE~ potential whose height is about $(0.003\,\eV)^4\sim\,$\Vde~ above the theoretically favored CC value of 0.\footnote{Our
idea depends on the vanishing CC assumption. At present, however, we do not
find any widely accepted CC solution. Any other ideas trying to explain DE also assume the vanishing CC somewhere, e.g. even in the quintessence idea of \cite{Steinhardt99} with the potential $\propto 1/\phi_{\rm quint}$ approaching to the current DE scale.
}
But then we go back to an extreme fine tuning problem. So, we propose to use suitable  discrete symmetries toward obtaining our desired approximate global symmetry \UDE.
Even before the 1998 discovery of the accelerating Universe, discrete symmetries were considered for obtaining some approximate global symmetry \cite{Garretson93}.  But with the 1998 discovery, the prospect of discrete symmetries as the basic tool for constructing global symmetries entered into the horizon of physics. Of course, the hypothetical discrete symmetry must satisfy the discrete gauge symmetry rule \cite{DiscrGauge89}.
Namely, we try to understand the DE scale from two {\it observed} fundamental mass parameters, the Planck mass $M_P$ and the TeV scale $v_{\rm ew}$, as envisioned in \cite{KimNilles13}.
Then, it has been argued that the suppression factor for the non-zero DE potential with the intermediate scale of $M_{\rm int}\simeq \sqrt{M_G v_{\rm ew}}$ is of order $1/M_G^7$ \cite{KimNilles13}. Since the GUT scale can be considered as the Planck scale multiplied by some O($\alpha\sim 10^{-2}$) parameters, we do not treat it as an independent scale.

The EW scale, $\mu$  and the TeV scale are of the same order and we treat them as one scale. The intermediate mass scale $M_I\sim F_a$ can be expressed as $\sqrt{M_G v_{\rm ew}}$ and we do not treat it as an independent scale. Indeed, the scale $\mu$ can be related to the intermediate scale in some models \cite{Kim13}. Since the DEPS~ scale is independent of the QCD scale, we will not consider it here. Summarizing, let us introduce the parameters related to the DEPS~ in terms of two scales,
\dis{
M_G~{\rm and}~ v_{\rm ew}, \label{eq:twoScales}
}
with the possibility of some O(1) couplings multiplied. If the DEPS~ potential of $\phiq$ is expressed in terms of $v_{\rm ew}$, $\phiq$ must couple to the pairs of Higgs doublets $H_u$ and $H_d$. If the combination $H_uH_d$ carries a PQ charge, $\phiq$ must have a PQ charge and we must worry about the anomaly \UDE--SU(3)$_c^2$. To remove this anomaly, we introduce another U(1) symmetry \UPQ~ so that one of the two linear combinations is free of the QCD anomaly. Namely, the DEPS~ scales (DEPS~ mass and decay constant) cannot be treated independently from the QCD axion scale.

Therefore, it is required to introduce two global U(1) symmetries to obtain the \UDE~ which is free of the \UDE--SU(3)$_c^2$ anomaly. As a consequence, two classes of singlet fields are introduced, $X$-type($X_i$ and $\OVER{X}_i$) for  the QCD axion and  $\Xq$-type($\Xq_i$ and $\OVER{\Xq}_i$) for the DEPS. If we introduce another confining force SU($N)_h$, an additional class of singlets will be needed not to introduce \UDE--SU(N)$_h^2$. In SUSY models, the intermediate axion scale $F_a\simeq 10^{10-12}\,\gev$ \cite{KSVZ} is considered as the confining scale of a hidden gauge group \cite{CKN92}. However, here we do not consider an extra nonabelian gauge group for simplicity and also because there may be some method to break SUSY without the extra nonabelian gauge group.

We attempt to obtain the DEPS~ of mass of order $10^{-32}\,\eV$ with the decay constant $\FDE~\gtrsim M_P$. The QCD axion has been discussed with the $S_2\times S_2$ symmetry \cite{Kim13}. Another discrete symmetry is introduced to house one more approximate global symmetry for the DEPS, $\Z_{N}$ or $\Z_{nR}$ \cite{KimDiscSymm13}. As a specific example, we consider the global symmetry $S_2\times S_2\times\Z_{10R}$.

The discussion is at the supersymmetric field theory level, even though it is strongly motivated by the orbifold compactification of the heterotic string. In Sec. \ref{sec:Demo}, we introduce the `quintessence-flavor democracy' to obtain the massless SM singlet fields, $\Xz, \Xzc, \Xqz,$ and $\Xqzc$.
In Sec. \ref{sec:DEpsV}, we introduce the breaking terms of global \UDE~ symmetry and obtain the tiny height of the DEPS~ potential, \Vde. The \UDE~ breaking  $A$-term dominates this breaking and renders the DEPS~ mass and the decay constant at  $m_{DEPS}\sim   10^{-32}\,\eV$ and $\FDE\gtrsim M_P$. Section \ref{sec:Conclusion} is a conclusion.

\section{Discrete symmetries}\label{sec:Demo}

The first step toward understanding the DE of the universe is to have a Goldstone field $\phiq^{(0)}$ where the superscript $^{(0)}$ means the Goldstone boson considering the lowest order interactions only. The next step is to make it a pseudo-Goldstone boson $\phiq$ of Fig. \ref{fig:Inflation} so that higher order contributions are feeble enough for the tiny vacuum energy of Fig. \ref{fig:Inflation}. So, we attempt to introduce an approximate global symmetry to have the pseudo-Goldstone boson. Since the global symmetry is approximate, this evades the gravity constraint that gravity does not respect global symmetries by the mother discrete symmetry \cite{GravityGlobal}. The mother discrete symmetry is the one respected by gravity \cite{DiscrGauge89, DiscreteR}. In this scheme, the needed global symmetries \UPQ~ and \UDE~ appear as accidental ones.

\begin{figure}[!t]
\begin{center}
\includegraphics[width=0.9\linewidth]{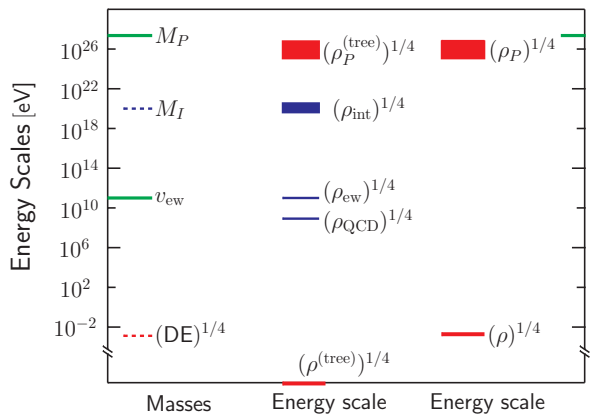}
\end{center}
\caption{Energy scales. The green color denotes the observed fundamental scales, $M_P$ and $v_{\rm ew}$. The red color denotes the DE related parameters. } \label{fig:MassScales}
\end{figure}

The Goldstone field is massless at the tree level ($\phiq ^{(0)}$), whose vacuum energy scale is depicted at the bottom red line in the second column of Fig. \ref{fig:MassScales}. On the other hand, the expected energy scale in the fundamental Lagrangian is  $(\rho_P^{(\rm tree)})^{1/4}$ which is depicted as the upper red band in the second column of Fig. \ref{fig:MassScales}. The key point is how to obtain the bottom red line $(\rho^{(\rm tree)})^{1/4}$ out of the GUT or the Planck scale energy $(\rho_P^{(\rm tree)})^{1/4}$. A similar question on the $\mu$ problem of SUSY has been answered by introducing `democracy' for two pairs of Higgs doublets \cite{KimDiscSymm13}. Here,  SUSY is not a necessary ingredient, but is useful to simplify the couplings.  So, for the `democracy' idea, fermions in terms of the left-handed(L) and the right-handed(R) fields are introduced \cite{DemoQuark} and we adopt SUSY for this purpose.

For a natural introduction of the DE scale, an extreme care is needed since any small correction at the GUT scale $M_G$, at the axion scale $F_a$, or even at the electroweak scale $v_{\rm ew}$, can easily outshoot the anticipated height of DEPS~ potential \Vde.

Let us introduce two chiral singlets $\Xq^{(1)}$ and $\Xq^{(2)}$ with SUSY and each with two chiralities: L and R chiralities. If these two singlets are not distinguished by any quantum number and geometry of the internal space, there must be the permutation symmetries, $S_2(L)$ and $S_2(R)$. The permutation symmetry $S_2$ is the permutation $1\leftrightarrow 2$, each for the L and R fields. The singlets, $ \Xq^{(1)}_{L}$ and $\Xq^{(2)}_{L}$ are the singlet representations of $S_2(L)$, and the singlets $\Xq^{(1)}_{R}$ and $\Xq_{R}^{(2)}$ (or represented by $\OVER{\Xq}_R^{\,(1)} \sim \Xq_{L}^{(1)c}$ and $\OVER{\Xq}_R^{\,(2)} \sim \Xq_{L}^{(2)c}$) are the singlet representations of $S_2(R)$. Then, the lowest order mass matrix $M_0$ of two singlets is $(\overline{\Xq}^{\,(1)},\, \overline{\Xq}^{\,(2)})_R M_0(\Xq^{(1)},\Xq^{(2)})^T_L$ where
\dis{
M_0= \left(\begin{array}{cc} \MGUT/2\,,\, &  \MGUT/2\\  \MGUT/2\,,\,  & \MGUT/2  \end{array}\right). \label{eq:demoMass2}
}
Here, $\MGUT$ is the GUT scale. Two eigenvalues of $M_0$ are $\MGUT$ and $0$, whose scales are schematically shown as $(\rho_P^{(\rm tree)})^{1/4}$ and  $(\rho^{(\rm tree)})^{1/4}$, respectively, in the second column of Fig. \ref{fig:MassScales}.
The mass matrix $M_0$ is diagonalized to
\dis{
 M^{\rm diag.}_0=\left(\begin{array}{cc} 0 & 0\\ 0&~~\MGUT
\end{array}\right), \label{eq:diagMassMatrix}
}
and the new bases are related to the old ones by
\dis{
\Xq^{(0),(G)}=\frac{1}{\sqrt2}\left(\Xq^{(1)}\mp \Xq^{(2)}\right),\label{eq:diagbases}
}
or
\dis{
\Xq^{(1),(2)}=\frac{1}{\sqrt2}\left(\mp \Xq^{(0)}+ \Xq^{(G)}\right).\label{eq:diagbases}
}
One may argue that starting with mass matrix (\ref{eq:diagMassMatrix}) is just postulating two mass eigenvalues from the outset. It is true but mass matrix (\ref{eq:diagMassMatrix}) must have an underlying symmetry, which will become transparent by the inverse transformation, \ie transforming Eq. (\ref{eq:diagMassMatrix}) to Eq. (\ref{eq:demoMass2}). A similar line of reasoning has been adopted toward a solution of the $\mu$ problem \cite{KimDiscSymm13}.

As stated in Eq. (\ref{eq:twoScales}), we attempt to express the DE scale in terms of the observed scales $M_P$ and $v_{\rm ew}$. Since $v_{\rm ew}$ is given by the vacuum expectation values(VEVs) of $H_u$ and $H_d$, $\phiq$ must couple to $H_uH_d$ to be expressed in terms of  $v_{\rm ew}$. Since $H_u H_d$ carries a PQ charge, $\phiq$ couples generically to the QCD anomaly. Thus, we need another pseudoscalar, say the QCD axion $a$, which also couples to $H_uH_d$ such that one linear combination of them can be constructed, not carrying the QCD anomaly. This anomaly free pseudoscalar is named again as DEPS~$\phiq$. In this vein, the criteria of Eq. (\ref{eq:twoScales}) connect the QCD axion and DEPS~ with two global U(1) symmetries: \UPQ~ and \UDE.

Since gravity does not respect global symmetries as commented in Ref. \cite{GravityGlobal}, we confine to discrete symmetries which are safe from gravity violation. Recently, simple criteria for obtaining $\Z_N$ and $\Z_{nR}$ from string compactification have been given \cite{KimDiscSymm13}. It relies on the VEVs of the SM singlet fields. To use this simple method, the full spectrum of massless particles should be given in the compactification process. In addition, due to two units of discrete charges of $H_u$ and $H_d$, there can result a $\Z_2$ parity which guarantees a $\Z_2$-odd WIMP particle.

In this paper, we introduce a discrete $Z_{nR}$ in addition to $S_2\times S_2$, to obtain the approximate \UDE~ and \UPQ. Here, $n$ must be large enough so that it does not to allow the same \UPQ~ and \UDE~ charges for two different type fields. Here, we choose $n=10$ so that the discrete symmetry is $S_2(L)\times S_2(R)\times \Z_{10R}$ for the MSSM fields plus heavy quarks and $X$-type and $\Xq$-type singlets. The permutation symmetry $S_2(L)$ is the permutation $1\leftrightarrow 2$ for $H_u^{(i)}$ or $X^{(i)}$ or $\Xq^{(i)}$ for $i=1,2$, and $S_2(R)$ is the permutation for $H_d^{(i)}$ or $\OVER{X}^{\,(i)}$ or $\OVER{\Xq}^{\,(i)}$. The relevant quantum numbers of the MSSM fields, heavy quarks and $X$-type and $\Xq$-type singlets are presented in Table \ref{tab:quintcharges}, including the electroweak(EW) hypercharge $Y_{\rm ew}$. In the lower two rows, we present two global charges of \UPQ~ and \UDE.

Motivated by the string compactification, we are interested in matter which belongs to ({\bf 248}, $\one$) and ($\one$, {\bf 248}) of E$_8\times$E$_8'$.  Since {\bf 248} is not a singlet under E$_8$, it cannot appear in $W$ as $W\sim M^2\,{\bf 248}$, \ie any  MSSM signet matter field cannot have a tadpole term in $W$. Any component of {\bf 248} can be classified by eight U(1) charges of the Cartan subalgebra of E$_8$. Any member, not belonging to the center of {\bf 248}, cannot have all the vanishing U(1) charges. Suppose a non-center member $\Xq$ of {\bf 248}. Certainly, not all the U(1) charges of $\Xq\Xq$ for a non-center $\Xq$ are vanishing, and $\Xq\Xq$ is not an E$_8$ singlet. On the other hand, there is a possibility that $\Xq\OVER{\Xq}$ can be an E$_8$ singlet. If $\Xq$ is  a center member, $\Xq\Xq$ term can be present. The above comment is checked for {\bf 3} of SO(3), $\three\cdot\three=(\psi^+\psi^-+\psi^-\psi^+)/2+\psi_3\psi_3$ where $\psi_3$ is a center member and $\psi^\pm$ are the non-center members. So, $\Xq\Xq$ in the superpotential is forbidden if $\Xq$ does not belong to the center.

For an SU($N$) subgroup of E$_8$, for example, the heavy quark $Q$ and $\Xq$ may belong to the same representation, and so may be the Higgs field and $X$,
\dis{
\left(\begin{array}{c} Q^{(r)}\\[0.2em] Q^{(g)}\\[0.2em] Q^{(b)}\\[0.3em] L^+\\[0.3em] L^0\\[0.2em] \Xq\\ \cdot\\ \cdot
\end{array}\right),~~
\left(\begin{array}{c} T^{(r)}\\[0.2em] T^{(g)}\\[0.2em] T^{(b)}\\[0.2em] H_u^+\\[0.4em] H_u^0\\[0.3em] X\\ \cdot\\ \cdot
\end{array}\right).\nonumber
}
The above grouping of fields is just an example.

But, here we introduce one different aspects for the $X$-type and $\Xq$-type fields. We allow
only one anomalous U(1) from the compactification. In this case, the anomalous gauge U(1)
is broken near the string scale and a global symmetry survives down to the low energy scale. Let this be the \UPQ~ because it carries the anomaly. So, the \UPQ~ breaking scale is at the intermediate
scale \cite{Kim88}. On the other hand, the \UDE~ is not anomalous and it must come from
the non-anomalous U(1) gauge symmetry. In principle, the spontaneous symmetry breaking scale
of this gauge U(1) is at the string scale. Then, the spontaneous symmetry breaking scale \UDE~
is determined by the VEVs of \UDE~ charge carrying singlet fields. Generically, these VEVs can
be raised to the Planck scale, since we do not use $B_{MN}$ but a phase field in matter fields as the DEPS.  Now, the  $X$-type and $\Xq$-type fields survive down to the low energy
scale where \UDE~ is already broken at $\gtrsim M_P$ and \UPQ~ is not broken.
Therefore, let us construct a model allowing the following features:
\begin{itemize}
\item Introduce $S_2(L)\times S_2(R)$. Introduce also a discrete symmetry $Z_{10R}$ which is large enough to introduce two approximate global symmetries, \UPQ~ and \UDE.

\item The \UPQ~ has the color anomaly, and the very light axion is housed in $X^{(0)}$ and $\OVER{X}^{\,(0)}$. Their VEVs  are at the intermediate scale.

\item The \UDE~ does not have the color anomaly, and the DEPS~ is housed in $\Xq^{(0)}$ and $\OVER{\Xq}^{\,(0)}$. For this purpose, it is necessary to have heavy quarks $Q_{(i)}$ and $\OVER{Q}_{\,(i)}$~($i=1,\cdots,N$).

\item There are three scales of VEVs. Out of two fundamental scales $M_G$ and $v_{\rm ew}$ of Eq. (\ref{eq:twoScales}), the intermediate scale $M_I$ and the DEPS~ parameters are derived.  $\Xq^{(0)}$ and $\OVER{\Xq}^{\,(0)}$ must couple to $H_u, H_d$  and to $N$ pairs of heavy quarks. We choose $N=2$.

\item The approximate global symmetries can be chosen into many directions. Among these,  \UPQ~ carries the color anomaly. For \UDE, the condition is the absence of \UDE--SU(3)$_c^2$ anomaly.
    We take the \UPQ~ breaking scale at the intermediate scale and the \UDE~ breaking scale at the Planck scale.

\item To have only one massless pairs through permutation symmetry, corresponding to the second column of Fig. \ref{fig:MassScales}, we introduce two pairs for each of $H_u-H_d$, $Q-\OVER{Q}$, $X-\OVER{X}$, and $\Xq-\OVER{\Xq}$.
\end{itemize}
\begin{widetext}
Therefore, the superpotential terms respecting the discrete symmetries of Table \ref{tab:quintcharges} are
\dis{
W  \propto &-M_{H^{(G)}}H_u^{(G)} H_d^{(G)}-M_{Q^{\,(I)}} Q^{(I)} \OVER{Q}^{\,(I)} -M_{\Xq^{(I)}} \Xq^{(I)} \OVER{\Xq}^{\,(I)}- M_{X^{(\kev)}} X^{(\kev)} \OVER{X}^{\,(\kev)}\\
&-f_{(u)}q u^cH_u^{(0)} -f_{(d)}q d^c H_d^{(0)}-\sum_{IJ}f_{IJ}u_I^c d_J^c \OVER{Q}^{\,(0)}
-\lambda_Q \Xq^{(0)}\left( Q^{(0)}\OVER{Q}^{\,(0)} + Q^{(0)}\OVER{Q}^{\,(I)} + Q^{(I)}\OVER{Q}^{\,(0)}  +\cdots\right)\\
&+\frac{\sum_{ij} X^{(0)} \OVER{X}^{\,(0)}}{M_P}(H_u^{(0)} H_d^{(0)}+\cdots)+ \frac{1}{M_P}\sum_{IJ}f_{IJ}u_I^c d_J^c (\OVER{Q}^{\,(0)}  \OVER{X}^{\,(0)}+ \cdots)
\\
&+\frac{\sum_{ij} \Xq^{(i)}  \OVER{\Xq}^{\,(j)}}{M_P^3}(H_u^{(0)} H_d^{(0)}+\cdots)^2 +\cdots \,.\\
    \label{eq:Wdiscrete}
}
Due to the $S_2\times S_2$ symmetry, the singlet mass $M_{H^{(G)}}$ is of order $M_G$ and one massless pair, $H_u^{(0)}$ and $H_d^{(0)}$, is obtained \cite{Kim13}. Without $S_2\times S_2$, the model can be more intricated to obtain the massless pair.  The masses  $M_{Q^{\,(I)}}$ and $M_{\Xq^{(I)}}$ are of order the intermediate scale due to the couplings $\langle\Xq\rangle\, \OVER{Q}^{(I)}Q^{(I)}$ and $\langle\Xq\rangle\, \OVER{\Xq}^{(I)}\Xq^{(I)}$. $ M_{X^{(\kev)}}$ is of order the keV scale due to the couplings $\langle \frac{H_uH_d}{M_G}\rangle\, \OVER{X}^{(\kev)}X^{(\kev)}$. The second line consists of dimension-3 $W$, the third line consists of dimension-4 $W$, the fourth line consists of dimension-6 $W$, etc., all of which satisfy the discrete symmetry $\Z_{10R}$ of Table \ref{tab:quintcharges}.

\begin{table}[!t]
\begin{center}
\begin{tabular}{|c|c|c|c|c|c|c|c|c|c|c|c|c|c|c|c|c|c|c|c|}
\hline &&&&& &&&&& &&&&& &&\\[-1.26em] Fields & $q_I$ &
$u_I^c$ & $d_I^c$ & $Q^{(i)}$ &$\OVER{Q}^{\,(i)}$ & $H_u^{(0)}$ & $H_d^{(0)}$& $H_u^{(G)}$ & $H_d^{(G)}$ & $X^{(0)}$ & $\OVER{X}^{\,(0)}$ & $X^{(G)}$ & $\OVER{X}^{\,(G)}$  & $\Xq^{(0)}$& $\OVER{\Xq}^{\,(0)} $& $\Xq^{(G)}$ & $\OVER{\Xq}^{\,(G)} $   \\
\hline\hline
$Y_{\rm ew}$  &$\frac16$ &$-\frac23$ &$+\frac13$& $-\frac13$ &$+\frac13$ & $+\frac12$& $-\frac12$& $+\frac12$& $-\frac12$ & $0$ &$0$ &$0$&$0$ &$0$&$0$&$0$&$0$ \\[0.3em]
\hline
$Z_{10R}$  &$+3$ &$+3$ &$+3$ & $+2$ & $+6$ & $+6$& $+6$& $+6$& $+6$ & $0$ &$0$ & $0$& $0$ &$+4$ & $+4$ &$+4$ &$+4$ \\[0.3em]
\hline
$\Gamma_{\rm PQ}$  &$0$ &$0$ &$+1$ & $-1$ &$ -1$ & $0$& $-1$ & $\times$ &$\times$&$+2$& $-1$& $\times$ & $\times$& $+2$ &$0$ & $\times$ &$\times$\\[0.3em]
$\Gamma_{\rm de}$  &$0$ &$+1$ &$0$& $-\delta$ &$-3+\delta$ & $-1$& $0$ & $\times$ &$\times$&$-1+\delta$&$+2- \delta$& $\times$ &$\times$  & $+3-\delta$ &$-1+\delta$ & $\times$ &$\times$\\[0.3em]
\hline
\end{tabular}
\end{center}
\caption{The PQ and quintessential charges from $S_2(L)\times S_2(R)\times Z_{10R}$. For the Higgs fields, $L$ refers to $H_u$ and $R$ refers to $H_d$. All the chiral fields are left-handed. The quantum numbers $\Gamma_{\rm PQ}$ and  $\Gamma_{\rm de}$ are given for the light fields with superscript $^{(0)}$. }\label{tab:quintcharges}
\end{table}
\end{widetext}

\begin{figure}[!t]
\begin{center}
\includegraphics[width=0.8\linewidth]{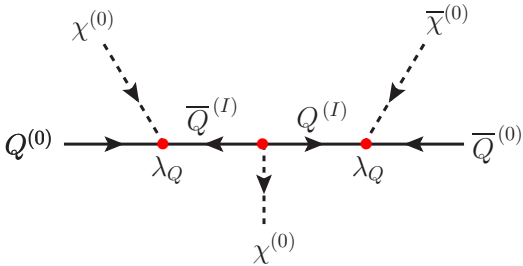}
\end{center}
\caption{Diagram raising the $Q^{(0)}-\OVER{Q}^{\,(0)} $ mass.
} \label{fig:HeavyQ00}
\end{figure}

Introduction of heavy quarks is necessary to have the DEPS~ together with the QCD axion since the SM quarks are assumed to carry the family-independent \UDE~ charges through the couplings to  $H_u^{(0)}$ and $H_d^{(0)}$. For the model to be viable with the gauge coupling unification, it is desirable to make all the $Q$-type quarks very heavy compared to the EW scale. With the interaction (\ref{eq:Wdiscrete}), indeed the massless pair at tree level, $Q^{(0)}$ and $\OVER{Q}^{\,(0)}$, obtains a huge mass due to the renormalizable $\lambda_Q$  term of Eq. (\ref{eq:Wdiscrete}), which is shown in Fig. \ref{fig:HeavyQ00}.
 The coefficients of supersymmetry condition $\partial W /{\partial Q}^{\,(0)}$ is zero,
\dis{
\frac{\partial W}{\partial Q^{(0)}} =0= \lambda_Q \Xq^{(0)} \OVER{Q}^{\,(0)} +\cdots.\label{eq:SUSYQQ}
}
Since $\langle \OVER{Q}^{\,(0)}\rangle=0$ not to break the color symmetry, there is no condition on $ \langle \lambda_Q \Xq^{(0)}  \rangle$ at this level.\\

Let us now proceed to discuss the approximate global symmetries \UPQ~ and \UDE.

The items 2) and 3) in Introduction are strong conditions. If we consider the SM gauge group below the GUT scale, the U(1) symmetries we can count are 12, left from the rank sixteen E$_8\times$E$_8'$ or SO(32) gauge symmetry. If the number of needed massless singlets at the third step for the independent U(1)$_{\rm global}$ charges is $n_1$, the $n_1$ global charge conditions cannot be met if $n_1>12$. For example, we need 11 independent charges of massless fields to define \UPQ~ and \UDE~  in Table \ref{tab:quintcharges}, and two global symmetries can be defined. If there is another confining force with rank $r_h$, for the conditions to be satisfied we require $n_1\le 12-r_h$.\footnote{Considering the anomalous U(1), this condition can be modified to $n_1\le 13-r_h$.} As an illustration, we present Appendix A.

\subsubsection{PQ symmetry}

We consider effective theory below the GUT scale. So, the PQ charges are those of superscript $^{(0)}$ fields of Table \ref{tab:quintcharges}. Here, we give an example that both $X$-type and $\Xq$-type fields couple to the SM quarks. In fact, it is not a strictly necessary condition at this field theory level discussion. If the PQ charges of the SM fermions are family dependent, introduction of heavy quarks may not be necessary. But,
if the PQ charges are family independent, it is necessary to introduce heavy quarks of Table \ref{tab:quintcharges} toward a global current which is free of the QCD anomaly. In this case, the heavy quark mass scale defines the PQ scale \cite{KSVZ}, and the chiral charges of the heavy quarks are the PQ charges. Next, from the last term of the second line of Eq. (\ref{eq:Wdiscrete}) the PQ charge of $\Xq^{(0)}$ is defined as $+2$.\footnote{We can also use $\OVER{\Xq}^{\,(0)}$ for this purpose.} Then, the third term in the second line of Eq. (\ref{eq:Wdiscrete}) fixes the sum of PQ charges of $u^c$ and $d^c$ as $+1$. We choose the PQ charges of $u^c$ and $d^c$ as $0$ and $+1$, respectively, which then fixes the PQ charge of the quark doublets $q$ as 0 from the first two terms in the second line of Eq. (\ref{eq:Wdiscrete}). Finally, the first term in the third line of Eq. (\ref{eq:Wdiscrete}) defines the sum of the PQ charges of $X^{\,(0)}$ and $\OVER{X}^{\,(0)}$ to be +1. We choose the PQ charges of $X^{\,(0)}$ and $\OVER{X}^{\,(0)}$ as +2 and $-1$, respectively. With these choices, the remaining two terms satisfy the PQ symmetry.
Note, however, that consideration of the GUT scale fields breaks the PQ symmetry \cite{Kim13}, which is the reason that we did not give their PQ charges in  Table \ref{tab:quintcharges}. This symmetry is indeed a PQ symmetry since the \UPQ-SU(3)$_c^2$ anomaly is proportional to
\dis{
\left(+1_{(d^c)}\cdot 3-1_{(Q^{(0)})}-1_{(\OVER{Q}^{\,(0)})}\right)  G_{\mu\nu}\tilde{G}^{\mu\nu}=G_{\mu\nu}
\tilde{G}^{\mu\nu},
}
and hence the axionic domain wall number is 1.

Thus, from Eq. (\ref{eq:Wdiscrete}), we define the PQ symmetry with $d=2,3,$ and 4 terms. The PQ symmetry breaking terms contain the terms of $d\ge 5$. The last term of Eq.   (\ref{eq:Wdiscrete}) belongs to this category and breaks the \UPQ~ symmetry.
\subsubsection{DE symmetry}

We must choose that the DEPS ~is orthogonal to the QCD axion. Let us start, for simplicity, the DE charge of $u^c$ is +1 and the DE charge of $d^c$ is 0.
A set of DE charges are shown in the last line of  Table \ref{tab:quintcharges}. There is no \UDE-SU(3)$_c^2$ anomaly,
\dis{
 \left(+1_{(u^c)}\cdot 3-\delta_{(Q^{(0)})}+(-3+\delta)_{(\OVER{Q}^{\,(0)})}\right)
 G_{\mu\nu}\tilde{G}^{\mu\nu}=0.
}

The \UDE--SU(2)$_W$--SU(2)$_W$  anomaly is not significant. Even if it is present, it is of order $M_Z^4\, e^{-2\pi\sin^2\theta_W/\alpha_{\rm em}} \simeq (M_Z\, e^{-49.5})^4\approx (10^{-11}\eV)^4$ which is much smaller than $(0.003\,\eV)^4$.

Let  $\langle X^{(0)}\rangle=\langle\OVER{X}^{\,(0)} \rangle=V_1$ and  $\langle\Xq^{(0)} \rangle=\langle\OVER{\Xq}^{\,(0)} \rangle=V_2$, with $V_2\gg V_1$. The Goldstone boson corresponding to the longitudinal degree of U(1)$_{\rm anom}$ gauge boson is
\dis{
a_{\rm anom}\propto V_1(2P_{X^{(0)}}-P_{\OVER{X}^{\,(0)}})+V_2(2P_{\Xq^{(0)}})\propto  P_{\Xq^{(0)}}.
\nonumber
}
The QCD axion $a$ has the same phase as that of $a_{\rm anom}$ but its decay constant is at the scale where the global \UPQ~ is broken at $\Mint$,
\dis{
a\propto V_1(2P_{X^{(0)}}-P_{\OVER{X}^{\,(0)}})+V_2(2P_{\Xq^{(0)}}).
}
The DEPS~ is proportional to,
\dis{
\phiq\propto &V_1\left((-1+\delta)P_{X^{(0)}}+(2-\delta)P_{\OVER{X}^{\,(0)}}\right)\\
&+V_2\left((3-\delta)P_{\Xq^{(0)}} +(-1+\delta)P_{\OVER{\Xq}^{\,(0)}}\right) .
}
The orthogonality of $a$ and $\phiq$ requires
\dis{
V_1^2\left[2(-1+\delta)-(2-\delta) \right]+V_2^2\left[2(3-\delta) \right]=0
}
so that $\delta$ is determined as a function of $V_1^2/V_2^2$. If $V_1^2/V_2^2=0$ is chosen, we obtain $\delta\simeq 3$. For $\delta=3$, the largest common divisor of the quantum numbers of $X^{(0)}, \OVER{X}^{\,(0)}, \Xq^{(0)}$ and $\OVER{\Xq}^{\,(0)}$ is 1. Thus, the domain wall number of the DEPS~ potential is 1.

\section{dark energy scale $(10^{-3}\,\eV)^4$}\label{sec:DEpsV}

Since the \Quint~ gives one pair of the quintessence singlets, \ie $\Xq^{(0)}$ and $\OVER{\Xq}^{\,(0)} $, zero mass as in the second column of Fig. \ref{fig:MassScales}, one has to break the \Quint~ to obtain a non-vanishing mass of $\Xq^{(0)}$ and $\OVER{\Xq}^{\,(0)} $. The \Quint~ must be broken for this purpose. We present an argument showing this possibility in a SUSY field theory framework.  Let us take the minimal K\"ahler potential $K=\Phi_i\Phi_i^\dagger$
where $\Phi_i\,(i=1,2)$ is the gauge group non-singlet field such as the Higgs superfield and $\Xq_i\,(i=1,2)$ are gauge group singlet superfields, obeying the common $S_2$ symmetry of $\Phi_i\,(i=1,2)$, both for L and R,\footnote{For the Higgs fields, L refers to $H_u$ and R refers to $H_d$.}
\dis{ S_2:
~\Phi_1 \leftrightarrow \Phi_2 ,~{\rm or}~ \Xq_1\leftrightarrow \Xq_2\,.\label{eq:CommonS2}
}
Let us consider the following $S_2(L)\times S_2(R)\times \Z_{10R}$ symmetric non-renormalizable terms, \dis{
W^{\rm (nonren.)} &= \tilde{\lambda}\sum_{i,j,k=1,2}\left(\frac{\Xq^{(i)} \OVER{\Xq}^{\,(i)} }{M_P^3}\right) H_u^{(j)}H_d^{(j)} H_u^{(k)}H_d^{(k)}\\
&+\tilde{\lambda}'\sum_{i,j,k=1,2}\left(\frac{\Xq^{(i)} \OVER{\Xq}^{\,(i)} }{M_P^3}\right) H_u^{(j)}H_d^{(k)} H_u^{(j)}H_d^{(k)}\\
&+\tilde{\lambda}''\sum_{i,j=1,2}\left(\frac{\Xq^{(i)}\OVER{\Xq}^{\,(j)} }{M_P^3}\right)H_u^{(i)}H_d^{(j)}H_u^{(k)}H_d^{(k)} \\
&+\cdots
\label{eq:KNterm}
}
which does not respect the PQ and DE symmetries given in Table \ref{tab:quintcharges}.
To have VEVs of $\Xq$ fields, let us consider an $S_2$ symmetric superpotential with a singlet $Z$ with $Z_{10R}$ quantum number 2 \cite{Kim84},
\dis{
W_{\rm int} &= Z \Big( X^{(1)} \OVER{X}^{\,(1)} +X^{(1)}\OVER{X}^{\,(2)}\\
 &+X^{(2)}\OVER{X}^{\,(1)} +X^{(2)}\OVER{X}^{\,(2)}  - \MI^2\Big)\,.\label{eq:WintChi} 
}

Let us concentrate on $X$ and $\OVER{X}$ of Eq. (\ref{eq:WintChi}) for a moment.
There exists a flavor-democracy breaking minimum, $\langle Z\rangle =0 , ~\langle X^{(1)}\rangle=\langle \OVER{X}^{\,(1)}  \rangle   = \MI , ~\langle X^{(2)}\rangle =\langle \OVER{X}^{\,(2)}  \rangle = 0$. Since there also exists the $S_2$ symmetric vacuum $\langle Z\rangle =0 , ~\langle X^{(1)}\rangle=\langle \OVER{X}^{\,(1)}  \rangle  =  ~\langle X^{(2)}\rangle =\langle \OVER{X}^{\,(2)}  \rangle \ne 0$, our choice of democracy breaking minimum is spontaneous.
The heavy fields are known to have very small VEVs. For example, we can expand the $X$-type fields around the VEVs, $\langle X^{(i)}\rangle $ and $\langle \OVER{X}^{\,(i)} \rangle $,
\dis{
  X^{(1)}&=\frac{\Mint}{2} +\frac{X}{2}-\frac{x}{2},~~\OVER{X}^{\,(1)}
  = \frac{\Mint}{2} +\frac{\OVER{X}}{2}-\frac{\bar x}{2},  \\
 X^{(2)} &=\frac{\Mint}{2} +\frac{X}{2}+\frac{x}{2},~~ \OVER{X}^{\,(2)}=\frac{\Mint}{2} +\frac{\OVER{X}}{2}+\frac{\bar x}{2},\label{eq:VACpert}
}
where $X=X^{(G)},~ x=X^{(0)},~ \OVER{X}= \OVER{X}^{\,(G)} $, and $\bar x= \OVER{X}^{\,(0)}$. Of course, the superpotential term $M_{X^{(G)}}X^{(G)}\OVER{X}^{\,(G)} $ does not contain the massless field $x$ and $\bar x$. Four complex scalars $X^{\,(1)},X^{\,(2)}, \OVER{X}^{\,(1)} $ and $\OVER{X}^{\,(2)} $ become four complex scalars $X, \overline{X}, x$ and $\bar x$ in Eq. (\ref{eq:VACpert}). For SUSY, the leading terms in the superpotential $M X\overline{X} + Z X \overline{X} $ require $(M+Z) X =0,~ (M+Z) \overline{X} =0$ and $X \overline{X}=0$. The heavy field solution $\langle X\rangle=\langle\overline{X} \rangle=0$ satisfy these conditions. But there is no condition on the light fields $x$ and $\bar x$. Their VEVs are determined when the flat directions along these fields are lifted when SUSY is broken. The VEVs $X^{(1)}=\Mint$ and $X^{(2)}=0$ fix  $x=\bar x=-\Mint/2$. How the magnitude of $x$ and $\bar x$ are chosen will be commented later, below Eq. (\ref{eq:DEpsV}).

At the democracy breaking vacuum, we mimic the above result by the following $W$ of $\Xq$ and $\OVER{\Xq}$
\dis{
W=M_{\Xq^{(G)}}\Xq^{\,(G)}\OVER{\Xq}^{\,(G)} +Z'\left(\frac{S'}{M_P}\Xq^{\,(0)}\OVER{\Xq}^{\,(0)} -{\cal M}^2\right),\label{eq:ZXXexample}
}
where $S'$ carries $Z_{10R}=4$ and $Z'$ carries $Z_{10R}=2$, and $M_{\Xq^{(G)}}\simeq \langle S'\rangle$.
At a SUSY vacuum, $\Xq^{\,(G)}=\OVER{\Xq}^{\,(G)} =Z'=0$ and $|\Xq^{\,(0)}|=|\OVER{\Xq}^{\,(0)} |= {\cal M}M_P/\langle S'\rangle$  can be chosen. Here, ${\cal M}$ is of order $M_P$. Therefore, the nonrenormalizable interaction becomes
\dis{
W &=(\tilde{\lambda} +\tilde{\lambda} '+\tilde{\lambda} ''+\cdots)\frac{\Xq^{(1)} \OVER{\Xq}^{\,(1)} }{M_P^3} H_u^{(1)}H_d^{\rm (1)}H_u^{(1)}H_d^{\rm (1)}\\
&=(\tilde{\lambda} +\tilde{\lambda}'+\tilde{\lambda} ''+\cdots)\frac{(\Xq^{(0)}+\Xq^{(G)})(\OVER{\Xq}^{\,(0)}+\OVER{\Xq}^{\,(G)} )}{8M_P^3}\\
&\quad\cdot (H_u^{(0)}+H_u^{\rm (G)})^2\, (H_d^{(0)}+H_d^{\rm (G)})^2\\
&\equiv \lambda\frac{( \Xq^{(0)} \OVER{\Xq}^{\,(0)}+\cdots )}{M_P^3} \Big(H_u^{(0)}H_d^{(0)}+H_u^{(0)}H_d^{(G)}\\
&\quad +H_u^{(G)}H_d^{(0)}+H_u^{(G)}H_d^{(G)} \Big)^2\\
&\longrightarrow \frac{\lambda }{M_P^3}\, \left(H_u^{(0)} H_d^{(0)}\right)^2\Xq^{(0)}\OVER{\Xq}^{\,(0)}
\label{eq:DEKN}
}
where $\lambda=(\tilde{\lambda}+\tilde{\lambda}'+\tilde{\lambda}''+\cdots )/8$ and in the last line heavy field VEVs are set to zero.

\begin{figure}[!t]
\begin{center}
\includegraphics[width=1\linewidth]{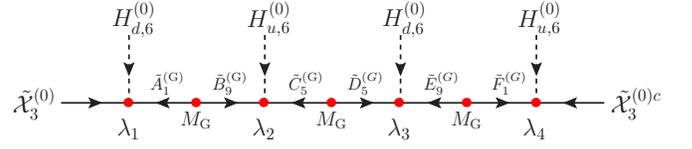}
\end{center}
\caption{The diagram defining \UDE~ charge. The subscripts are the $Z_{10R}$ quantum numbers, and  $\Xq^{(0)c}$ is $\OVER{\Xq}^{\,(0)}$.
} \label{fig:DEqnDef}
\end{figure}

The defining diagram for the PQ and DE charges from the $Z_{10R}$ symmetry is shown in Fig. \ref{fig:DEqnDef}. The subscripts are the
$Z_{10R}$ quantum numbers. At each vertex, the  $Z_{10R}$ symmetry is preserved. The effective interaction is the last line of Eq. (\ref{eq:DEKN}), and the PQ and DE quantum numbers of the light fields, $H_u^{(0)}, H_u^{(0)}, \Xq^{(0)}$, and $\OVER{\Xq}^{\,(0)}$ are those given in Table \ref{tab:quintcharges}. Assuming that the GUT scale masses of the intermediate lines of Fig. \ref{fig:DEqnDef} are the same, the parameter $\lambda$ of Eq. (\ref{eq:DEKN}) is expressed in terms of parameters of Fig. \ref{fig:DEqnDef} as
\dis{
\lambda=\frac{\lambda_1 \lambda_2\lambda_3\lambda_4 M_P^3}{8M_G^3}.
}

\begin{figure}[t!]
\begin{center}
\includegraphics[width=0.95\linewidth]{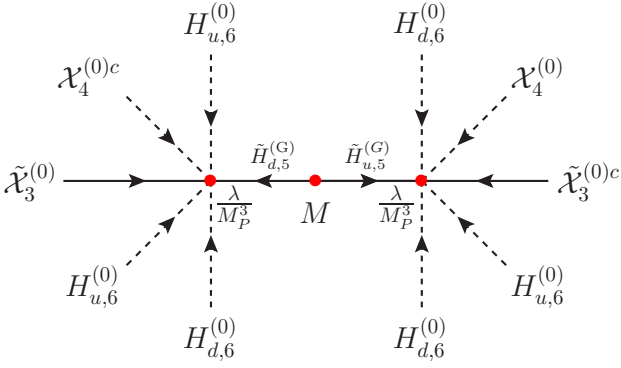}
\end{center}
\caption{The leading \UDE~ violating diagram with $\Xq^{(0)c}$ representing $\OVER{\Xq}^{\,(0)}$. Here, the $Z_{10R}$ charges are shown as the subscipts, which is an explicit realization of the idea presented in Fig. 3 of Ref. \cite{KimNilles13}.
} \label{fig:TwoDEpsino}
\end{figure}

The effective theory in terms of the light fields only (with superscript $^{(0)}$) has the following features. Firstly, the mass term is vanishing if the symmetry $S_2\times S_2$ is unbroken. At the $S_2\times S_2$ breaking vacuum, however, the mass term is generated. Second, the dimensionless couplings get renormalized,  dominantly by the logarithmic evolution. If SUSY is assumed, then the Yukawa couplings are not renormalized. The magnitudes of the dimensionless couplings are fixed if the theory is given.  Third, the heavy particles generate the nonrenormalizable interactions of the light fields.

The third line of Eq. (\ref{eq:DEKN}) is the one to study the SUSY conditions,
\dis{
 \frac{\lambda}{M_P^3} &\Big(H_u^{(0)}H_d^{(0)}+H_u^{(0)}H_d^{(G)}+H_u^{(G)}H_d^{(0)}\\
 & +H_u^{(G)}H_d^{(G)} \Big)^2 (\Xq^{(0)}\OVER{\Xq}^{\,(0)}+\cdots ).\label{eq:SUSYcondChi}
}
The dominant term is made zero by the condition $\partial W/\partial H_i^{(0)}$,
\dis{
\frac{\partial W}{\partial H_u^{(0)}}\longrightarrow& \frac{\lambda}{M_P^3}(H_d^{(0)}+H_d^{(G)})( H_u^{(0)}H_d^{(0)}+\cdots)\\
&\cdot\Xq^{(0)}\OVER{\Xq}^{\,(0)}= 0,~\cdots\label{eq:tevderi}
}
\dis{
\frac{\partial W}{\partial H_u^{(G)}}\longrightarrow &\frac{\lambda}{M_P^3}(H_d^{(0)}+H_d^{(G)})( H_u^{(0)}H_d^{(0)}+\cdots)\\
&\cdot \Xq^{(0)}\OVER{\Xq}^{\,(0)} +M_{G}H_d^{(G)}= 0,~\cdots  \label{eq:GUTderi}
}
From Eq. (\ref{eq:GUTderi}), we obtain almost vanishing VEVs for the GUT scale fields,
\dis{
\langle H_d^{(G)}\rangle\simeq 0  ,~~\langle H_u^{(G)}\rangle\simeq 0  ,\label{eq:HVEVrelations}
}
and similarly
\dis{
\langle \Xq^{(G)}\rangle\simeq 0 ,~~\langle \OVER{\Xq}^{\,(G)}\rangle\simeq 0  .\label{eq:ChiVEVrelations}
}
With these choices, the derivative equations of superscript $^{(0)}$ fields such as Eq. (\ref{eq:tevderi}) seem not satisfied, but Eq.
(\ref{eq:tevderi}) is a highly suppressed term compared to the leading term of Eq. (\ref{eq:GUTderi}).

On the other hand, the light fields have flat directions which are lifted by including  soft terms, leading to the intermediate scale VEVs of $\Xq^{(0)}$ and $\OVER{\Xq}^{\,(0)}$,
\dis{
&\langle \Xq^{(0)}\rangle\simeq M_{\rm int},~\langle \Xq^{(G)}\rangle\simeq 0 ,\\
&\langle \OVER{\Xq}^{\,(0)}\rangle\simeq   M_{\rm int},~\langle \OVER{\Xq}^{\,(G)}\rangle\simeq 0  \,.\label{eq:ChiVEVrela}
}
As commented just above the bullet items in Sec. \ref{sec:Demo}, the anomaly-free \UDE~ is spontaneously broken at $\gtrsim M_P$. The additional \UDE~ breaking VEVs do not change this scale very much since the contribution of VEVs in Eq. (\ref{eq:ChiVEVrela}) to the \UDE~ breaking scale is $\gtrsim \sqrt{M_P^2+aM_{\rm int}^2}$ where $a$ is determined by the charges of the \UDE~ breaking scalar fields.

Summarizing, we expect the following gauge invariant superpotential terms involving $\Xq^{(0)}$ and $\OVER{\Xq}^{\,(0)}$,
\dis{
\varepsilon  &\Xq^{(0)}\OVER{\Xq}^{\,(0)}+\lambda_1 H_u^{(0)} H_d^{(0)} \Xq^{(0)} +\lambda_2 H_u^{(0)} H_d^{(0)} \OVER{\Xq}^{\,(0)}\\
&+ \frac{f}{M_P } \left(H_u^{(0)} H_d^{(0)}\right) \Xq^{(0)}\OVER{\Xq}^{\,(0)}\\
&+\frac{\lambda}{M_P^3} \left(H_u^{(0)} H_d^{(0)}\right)^2 \Xq^{(0)}\OVER{\Xq}^{\,(0)}+\cdots.
\nonumber
}
where the last term is given in Fig. \ref{fig:DEqnDef}.
When the VEVs of GUT scale fields are inserted, the light field interactions include more terms. If $H_{u,d}^{(G)}$ develop VEVs, then the light field interactions would include SU(2)$\times$U(1) breaking terms too. In the effective theory in the sense of Wilson, thus we consider the following interactions of light fields,
\dis{
W &=\mu H_u^{(0)}H_d^{(0)}+ \varepsilon \Xq^{(0)}\OVER{\Xq}^{\,(0)}+ a  \Xq^{(0)} H_u^{(0)}+ b\OVER{\Xq}^{\,(0)} H_d^{(0)}\\
& + a'  \Xq^{(0)} H_d^{(0)}+ b'\OVER{\Xq}^{\,(0)} H_u^{(0)} +\lambda_1 H_u^{(0)} H_d^{(0)} \Xq^{(0)} \\
& +\lambda_2 H_u^{(0)} H_d^{(0)} \OVER{\Xq}^{\,(0)}+ \frac{f}{M_P}H_u^{(0)} H_d^{(0)} \Xq^{(0)}\OVER{\Xq}^{\,(0)}\\
& +\frac{\lambda}{M_P^3} \left(H_u^{(0)} H_d^{(0)}\right)^2 \Xq^{(0)}\OVER{\Xq}^{\,(0)}+\cdots.
\label{eq:effCoups}
}
\vskip 0.3cm

We proceed to estimate the DEPS~ mass by considering the \UDE~ breaking:\\

\noindent {\it (a) From non-renormalizable interactions}--
Firstly, let us look at the non-renormalizable $\lambda$ term of Eq. (\ref{eq:effCoups}). With SUSY, the pseudo-Goldstino mass is given by the last line of Eq. (\ref{eq:DEKN}). Namely,  Fig. \ref{fig:DEqnDef} does not break the \UDE~ symmetry of Table \ref{tab:quintcharges}. The Goldstone boson DEPS~ obtains mass by Fig. \ref{fig:TwoDEpsino} since it breaks the \UDE~ charge of Table \ref{tab:quintcharges} by two units. Let us look at the SUSY and soft terms which break the \UDE~ symmetry. It turns out that the soft term dominates.

The superpotential obtained from Figs. \ref{fig:DEqnDef} and \ref{fig:TwoDEpsino} is
\dis{
W = &\frac{\lambda }{ M_P^3}H_u^{(0)} H_d^{(0)}H_u^{(0)} H_d^{(0)}\Xq^{(0)}\OVER{\Xq}^{\,(0)}\\
&+\frac{\lambda^2 \xi}{ M_P^6 M_G}\left[H_u^{(0)} H_d^{(0)}\right]^3 \left[\Xq^{(0)} \OVER{\Xq}^{\,(0)}\right]^2,
\label{eq:W}
}
where $M=M_G/\xi$. Let $\FDE=\eta M_P$.
Then, the last term of (\ref{eq:W}) breaking \UDE~ is
\dis{
\Delta W  &= \frac{\lambda^2 \xi}{M_P^2 M_G}\left|\frac{\Xq^{(0)} \OVER{\Xq}^{\,(0)}}{M_P^2} \right|^2  \left|\frac{ v_uv_d}{2}\right|^3\\
 &=\frac{\lambda^2\xi v^6_{\rm ew}}{8M_P^2 M_G}\left|\frac{\Xq^{(0)} \OVER{\Xq}^{\,(0)}}{M_P^2} \right|^2  \left|\frac{\tan\beta }{1+\tan^2\beta} \right|^3\\
 &= \frac{\lambda^2\xi\eta^4 v^6_{\rm ew}}{8M_P^2 M_G}   \left|\frac{\tan\beta }{1+\tan^2\beta} \right|^3\\
&=1.9\times 10^{-40}\lambda^2\xi\eta^4 \left|\frac{\tan\beta }{1+\tan^2\beta} \right|^3   [\gev^3], \label{eq:DEpsW}
}
So the SUSY breaking $A$-term is given by
\dis{
\Delta V  =3.8\times 10^{-40}\lambda^2\xi\eta^4 \left|\frac{\tan\beta }{1+\tan^2\beta} \right|^3
\left(\frac{m_{3/2}}{\gev}\right)\,[\gev^4]. \label{eq:DEpsV}
}
For the height of $V$ in the range of $10^{-46}[\gev]^4$ in the large $\tan\beta$ limit, say $\tan\beta=40$, we need $ 6\times 10^{-42}\lambda^2\xi\eta^4 m_{3/2 ,\tev}\approx 10^{-46}$, or $\lambda^2\xi\eta^4 m_{3/2 ,\tev}\gtrsim 1.7\times 10^{-5}$. The inequality is needed so that a smaller misalignment angle may fit to give the needed DE scale \Vde.

\vskip 0.5cm
\noindent {\it (b) From renormalizable interactions}-- Second, we consider the mass terms and masses generated by the renormalizable interactions. For $a,b,a'$, and $b'$ of Eq. (\ref{eq:effCoups}), we consider the following $4\times 4$ ino mass matrix $\overline{\Psi}\, m_{\rm ino}\Psi$
\dis{
m_{\rm ino}=\frac12\left(\begin{array}{cccc} 0& \varepsilon & a& a'\\
\varepsilon  &0 & b' & b\\ a& b' & 0 & \mu \\ a' & b& \mu &0
\end{array} \right)\label{eq:inomassmatrix}
}
where
\dis{
\overline{\Psi}=(\Xq^{(0)}, \OVER{\Xq}^{\,(0)},\, H_u^{(0)},\, H_d^{(0)}), ~
\Psi=\left(\begin{array}{c} \Xq^{(0)} \\[0.3em] \OVER{\Xq}^{\,(0)}\\[0.3em]
H_u^{(0)} \\[0.3em] H_d^{(0)}  \end{array} \right).
}
The parameters in Eq. (\ref{eq:inomassmatrix}) are given
below the electroweak scale, satisfying the SU(3)$_c\times$U(1)$_{\rm em}$ symmetry.
The value $(\Delta m_q)^{\rm re} $ is estimated from Fig. \ref{fig:DEqnDef} (Eq. (\ref{eq:DEKN})),
\dis{
(\Delta & m_q)^{\rm re} \simeq  \lambda\langle H_u^{(0)} \rangle^2 \langle H_d^{(0)} \rangle^2 /M_P^3\\
&= \lambda  v_{\rm ew}^4 \tan^2\beta\cos^4\beta /M_P^3= 2.6\times 10^{-39}\lambda\,[\eV]\\
 \label{eq:etsvarepsilon}
}
where $\tan\beta\approx 10$ is used in the last line.

To estimate $a,b,a'$ and $b'$, note that the VEVs with superscript $^{(G)}$ is highly suppressed.
Consider the order of magnitude of $a,a',b,b'$,
\dis{
a, a', &b,b' \sim \frac{\lambda}{\Mp^3} (\VEV{\Xq^{(G)}}\VEV{H^{(0)}}^3+\VEV{\Xq^{(0)}}\VEV{H^{(G)}}^3 )\\
&\sim \lambda\left(\frac{v_{\rm ew}}{\Mp}\right)^3 v_{\rm ew}\sim 2.5\times 10^{-37}\lambda\,[\eV] ,
\label{eq:ababprimes}
}
where $ \tan\beta\approx 10$ is used in the last line.
The determinant  of $2m_{\rm ino}$ is $(ab-a'b')^2 -2(ab+a'b')\varepsilon \mu
+ \varepsilon^2 \mu^2$. Therefore, in our parameter range the determinant is dominated by the term $\varepsilon^2\mu^2$, and hence an additional contribution to the DEPS~ mass is $\varepsilon$, Eq. (\ref{eq:effCoups}).

\section{Conclusion}\label{sec:Conclusion}

We introduced two discrete symmetries $S_2\times S_2$ and $Z_{10R}$ to obtain two approximate global symmetries \UPQ~ and \UDE. From the flavor-democracy among the $X$- and $\Xq$-type fields, we first obtain massless particles $X^{(0)}, \OVER{X}^{\,(0)}, \Xq^{\,(0)}$, and $\OVER{\Xq}^{\,(0)}$. In some string models, there appear two pairs of particles from which one massless pair can be easily obtained. When two approximate global symmetries are spontaneously broken, the corresponding Goldstone bosons (the very light axion and the DEPS) obtain masses. The QCD anomaly gives the mass to the axion, and the color-anomaly free DEPS~ mass is smaller than the axion mass. To set the DEPS ~potential at the DE scale, one must obtain a very tiny DEPS~ mass.  With the SUSY extension, we show a possibility of the needed DEPS~ potential height of order \Vde~ dominated by the SUGRA $A$-term and the  DEPS~ decay constant $\FDE$ in the range of Planck scale. Then, the DEPS~ mass falls in the region $\sim 10^{-32}\,\eV$. In our discussion, the discrete symmetry $S_2(L)\times S_2(R)\times \Z_{10R}$ is the underlying symmetry respected by gravity, and hence this mechanism evades the old gravity dilemma on global symmetries.

\section*{Acknowledgments}
This work is supported in part by the National Research Foundation (NRF) grant funded by the Korean Government (MEST) (No. 2005-0093841) and by the IBS(IBS CA1310).

\vskip 0.5cm

\begin{appendix}
\section{Approximate global symmetry from string}

Let us consider the $qd^c H_d^{(0)}$ term in the second line of Eq. (\ref{eq:Wdiscrete}). In the heterotic string, four U(1)'s beyond the rank-4 SM group from E$_8$ are denoted as
\dis{
 &Q_1=(1\,1\,1\,1\,1\,0\,0\,0)\,(0^8)',\\
  &Q_2=(0\,0\,0\,0\,0 \, 1\,0\,0)\,(0^8)',\\
 &Q_3= (0\,0\,0\,0\,0\,0\,1 \,0)\,(0^8)',\\
 &Q_4= (0\,0\,0\,0\,0 \,0\,0\,1)\,(0^8)',
}
and four U(1)'s from E$_8'$ are denoted as
\dis{
 &Q_5= (0^8)\,(1\,1\,1\,1\,1\,0\,0\,0)',\\
  &Q_6= (0^8)\,(0\,0\,0\,0\,0 \, 1\,0\,0)',\\
 &Q_7= (0^8)\, (0\,0\,0\,0\,0\,0\,1 \,0)' ,\\
 &Q_8= (0^8)\, (0\,0\,0\,0\,0 \,0\,0\,1)',\\
 &Q_9= (0^8)\, (1\,-1\,0\,0\,0 \,0\,0\,0)',\\
 &Q_{10}= (0^8)\, (1\,1\,-2\,0\,0 \,0\,0\,0)',\\
 &Q_{11}= (0^8)\, (1\,1\,1\,-3\,0 \,0\,0\,0)',\\
 &Q_{12}= (0^8)\, (1\,1\,1\,1\,-4 \,0\,0\,0)'.
 }
Let the $\Z_{10}$ is derived by the VEVs of $s_9$ and $s_{13}$ of Table \ref{tab:DoubleSU5} as discussed in \cite{KimDiscSymm13}. This $\Z_{10}$ is a discrete subgroup of U(1)$_{10}$ whose generator is a linear combination of $Q_i\,(i=1,2,\cdots,8)$,
\dis{
Z_{10}=\sum_{i=1}^8 \alpha_i Q_i.
}
The breaking condition, U(1)$_{10}\to\Z_{10}$ by $\langle s_9\rangle=\langle s_{13}\rangle$, is
\dis{
-\frac12(\alpha_2+\alpha_3+\alpha_4)+\frac34\alpha_6-\frac14\alpha_7-\frac12\alpha_8=0.
}
The fields $q, d^c,$ and $H_d$ are also given in Table \ref{tab:DoubleSU5}. Note that the condition for the existence of  $q d^c H_d$ term from the U(1)'s from E$_8\times$E$_8'$ is automatically satisfied.

\begin{table}[!t]
\begin{center}
\begin{tabular}{|c|c|c|c|  }
\hline &&&  \\[-1.26em]
Sector&  $n\cdot$Field &  Weight & $Z_{10}$   \\
\hline\hline
$T_4^{(0)}$ &$2\cdot \, H_d$ &$(0\,0\,0\,\underline{1\,0}\, \frac13\,\frac13\,\frac13\,)(0^8)'$ &$ +2$  \\[0.3em]
$T_4^{(0)}$ &$2\cdot \, H_u$ &$(0\,0\,0\,\underline{-1\,0}\,  \frac13\,\frac13\,\frac13\,)(0^8)'$ &$ +2$  \\[0.3em]
$T_4^{(0)}$ &$2\cdot\, q$ &$(\underline{\frac12\,\frac12\,\frac{-1}{2}}~\underline{\frac12\,\frac{-1}{2}}\, \frac{-1}{6}\,\frac{-1}{6}\,\frac{-1}{6}\,)(0^8)'$ &$ -1$  \\[0.3em]
$T_4^{(0)}$ &$2\cdot\, d^c$ &$(\underline{\frac12\,\frac{-1}{2}\, \frac{-1}{2}}~  \frac{-1}{2}\, \frac{-1}{2}\, \frac{-1}{6}\,\frac{-1}{6}\,\frac{-1}{6}\,)(0^8)'$ &$ -1$  \\[0.3em]
$T_3$ &$ s_9$ &$( 0^5~\frac{-1}{2}\,\frac{-1}{2}\, \frac{-1}{2})(0^5~\frac{3}{4}\,\frac{-1}{4}\,\frac{-1}{2})'$ &$ -10$  \\[0.3em]
$T_9$ &$ s_{13}$ &$( 0^5~\frac{1}{2}\,\frac{1}{2}\, \frac{1}{2})(0^5~\frac{-3}{4}\,\frac{1}{4}\,\frac{1}{2})'$ &$ +10$  \\[0.3em]
$T_9$ &$2\cdot\,  s_{14}\equiv X$ &$( 0^5~\frac{1}{2}\,\frac{-1}{2}\, \frac{-1}{2})(0^5~\frac{-3}{4}\,\frac{1}{4}\,\frac{1}{2})'$ &$ +8$  \\[0.3em]
$T_9$ &$2\cdot\,  s_{15}\equiv \OVER{X}$ &$( 0^5~\frac{-1}{2}\,\frac{-1}{2}\, \frac{1}{2})(0^5~\frac{1}{4}\,\frac{-3}{4}\,\frac{1}{2})'$ &$-10$  \\[0.3em]
\hline
\end{tabular}
\end{center}
\caption{The $Z_{10}$ quantum numbers for some fields of Ref. \cite{HuhKK09}, calculated in \cite{KimDiscSymm13}. These $Z_{10R}$ charges are different from those of Table \ref{tab:quintcharges}.  }\label{tab:DoubleSU5}
\end{table}

If all the terms in Eq. (\ref{eq:Wdiscrete}) are satisfied as the $qd^cH_d$ term does, then we are discussing a gauge symmetry. The global charges cannot be automatically satisfied in this way. To check the global symmetry we also introduced $X$ and $\OVER{X}$  in
Table \ref{tab:DoubleSU5}. Under $\Z_{10R}$, the term $H_uH_d X\OVER{X}$ can be present since it carries $+2$ units of $Z_{10}$ charge. For this term to be present from the U(1)$_{10}$ gauge symmetry, we have another conditions on the $Z_{10R}$ charges of $X$ and $\OVER{X}$ of Table \ref{tab:quintcharges},
\dis{
 &\frac12\alpha_2-\frac12(\alpha_3+\alpha_4) -\frac34\alpha_6 +\frac14\alpha_7+\frac12\alpha_8=0,\\
 & -\frac12(\alpha_2+\alpha_3+\alpha_4) +\frac14\alpha_6 -\frac34\alpha_7+\frac12\alpha_8=0.
}
Because there are nonzero $Z_{10R}$ charges, the coefficients $\alpha_i$ are completely determined by twelve charge conditions if E$_8'$ does not lead to a nonabelian gauge group.

\end{appendix}


\end{document}